\documentstyle[aps,prl,epsfig,twocolumn]{revtex} 
\begin{document}
\twocolumn[\hsize\textwidth\columnwidth\hsize\csname
@twocolumnfalse\endcsname

\draft
\title{The Cone Phase of Liquid Crystals:
Triangular Lattice of Double-Tilt Cylinders}

\author{Yashodhan Hatwalne  and  N.  V.
Madhusudana} \address{ Raman Research Institute,
Bangalore - 560 080, INDIA}

\date{\today}

\maketitle 
\widetext
\begin{abstract} 
We predict the existence of a new  defect-lattice phase near
the nematic - smectic-$ C $ ($ NC $) transition. 
This tilt- analogue of the blue phase is a  lattice
of {\em double-tilt cylinders}. We discuss the structure and 
stability of the cone phase. We suggest that many
`nematics' exhibiting short range layering and tilt order may in
fact be in the molten cone phase, which is a line liquid.
\end{abstract}

 ]
\narrowtext

In the nematic to smectic-$ C $ ($ NC $) transition the continuous 
translational symmetry of the nematic is spontaneously broken and a layered 
structure with unit layer normal $ {\bf N} $ develops
~\cite{deGProst}. In addition, smectic-$C$ liquid crystals have 
{\em tilt order} which is characterized by the projection $ {\bf c} $ 
of the apolar nematic director $ {\bf n}  $ onto the smectic layers. 
The smectic-$C$ structure belongs to the symmetry group $ C_{2h} $ 
(a mirror plane, a two-fold rotation axis normal to the mirror plane, 
and a centre of inversion). Note in particular that this biaxial structure
{\em spontaneously breaks the continuous azimuthal symmetry}; the average 
orientation of the $ {\bf c} $ vector is fixed in space (Fig. ~\ref{smc}).

In this letter we address the following question: Can a phase in which 
the smectic layer normal explores the entire range ($ 0 $ to $ 2 \pi $) of 
azimuthal angles about the nematic director $ {\bf n} $ exist?  We 
answer this question in the affirmative, propose a possible defect-
lattice structure (Fig.~\ref{conephase}) and  analyse the mechanism
which stabilizes it. Liquid crystals are {\em soft} materials; this makes
the existence of well-known defect-lattice phases such as the blue 
phases and the twist grain boundary phases ~\cite{deGProst} possible. 
Our discussion has certain similarities with the {\em low chirality}
stability analysis of the blue phases ~\cite{deGProst,MSAB}. It is
therefore useful to draw an analogy between the stabilizing mechanisms of
the blue phases and the cone phase. 
The cholesteric phase (which has a single twist axis) is unstable to 
the formation of the blue phases (which are lattices made up of 
double-twist cylinders) if the coefficient of the saddle-splay term in 
the molecular director $ {\bf n} $ in the Frank free energy is {\em
positive}. The blue phases satisfy
the tendency of chiral molecules to sustain twist in all possible directions.
The smectic-$C$ phase (which has tilt order in one direction) is unstable to 
the formation of the cone phase (which is a defect lattice made up of
tilt disclinations) if the coefficient of the Gaussian curvature term 
(which is the saddle-splay- like term in the smectic layer normal 
$ {\bf N} $) in the free energy is {\em negative}.  The cone phase is a
lattice of double-tilt cylinders in which the smectic layer normal tilts
in all possible directions about the nematic director.
If the coefficient of the Gaussian curvature term in the free energy
is {\em positive} in a smectic with tilt order, the cone phase is
clearly not stabilized. Instead, we expect `plumber's nightmare' phases
of the type discussed in ~\cite{Levelut}.  

Our principal results are  as follows: We predict the existence of
the cone phase of liquid crystals near the first order $ NC $ transition
and discuss the mechanism responsible for stabilizing it.
The cone phase is a triangular lattice of double-tilt cylinders, each
of which is a +1 disclination (in both the $ {\bf c} $- field as well as
the $ {\bf N} $- field) in the form of a stack of conical
layers. For reasonable values of parameters we estimate
that the cone phase composed of double-tilt cylinders of radius $ 50 \AA $ 
would be stabilized if the coefficient of the Gaussian curvature term in
the free energy $ \kappa_{G} \simeq - 3 \times 10^{-6} $ dyne. Based on
our theoretical analysis and the results of X-ray and neutron diffraction
experiments on many compounds exhibiting {\it skew-cybotactic} order, we
surmise that this so-called `nematic' phase with short-range layering/tilt
order may in fact be the molten cone phase - a line liquid of double-tilt
cylinders.  

The de Gennes model ~\cite{deGennes} of the $ NC $ transition
and the Chen-Lubensky $ NAC $ model ~\cite{CL} are motivated by the 
experimental observation that x-ray scattering shows two rings peaked around 
$ q_{c} = ( q_{\perp} \cos \phi, q_{\perp} \sin \phi, \pm q_{\|} ) $
in the vicinity of the $ NC $ transition. {\em Pretransitional fluctuations 
clearly explore all the azimuthal angles}. Fluctuations drive the
second- order mean field $ NC $ transition of the Chen-Lubensky model
to a first order transition ~\cite{B,S}. The deGennes model as
well as the Chen-Lubensky model  predict that all the Frank elastic
constants diverge at the $ NC $ transition. This has been borne out 
experimentally ~\cite{WHH}. 
  
Clearly, any distortion in the nematic director is enormously costly
near the $ NC $ transition. However, it is possible to have   
stable configurations where the smectic layer
normal exhausts the full range of azimuthal angles relative to a
{\em distortion-free} nematic director. Such field configurations are
topologically nontrivial and necessarily involve disclination lines
(Fig.~\ref{cylinder}). The cone configuration has a positive, 
delta-function Gaussian curvature along its axis, which screens the 
disclination charge ~\cite{NP,DN,PL}. It is therefore essential to use
covariant elasticity to calculate the energetics of the disclination
lines. Our formulation naturally incorporates the elasticity theory of 
membranes with tangent-plane order.  
     
For the stability of the cone phase, it is essential that the $ NC $
transition be first order (see the discussion following equation 
(\ref{eqnarray:conenergy})). This is the case for almost all compounds 
exhibiting the $ NC $ transition. To  describe this transition it is
sufficient for our purpose to use a simple phenomenological 
free energy density 
\begin{equation}
\label{eq:fL}
f_{L} = a (T_{c} - T),
\end{equation}  
where $ a $ is a constant. The free 
energy density $ f_{L} $ accounts for the condensation energy of the
smectic layering as well as that for the development of tilt order at
the $ NC $ transition.

The Frank free energy density 
\begin{equation}
\label{eq:fn}
f_{n}   =  \frac{K_{1}}{2} (\nabla \cdot {\bf n})^{2} 
           + \frac{K_{2}}{2} ({\bf n} \cdot \nabla \times {\bf n})^{2} 
+ \frac{K_{3}}{2} ({\bf n} \times \nabla \times {\bf n})^{2} 
\end{equation}
describes the energy cost for distortion in the Frank director
~\cite{deGProst}. In 
(\ref{eq:fn}) $ K_{1}, K_{2} $ and $ K_{3} $ are respectively the splay,
twist and bend elastic constants.

The free energy density for distortions in the smectic layering is 
given by
\begin{equation}
\label{eqn:fs}
f_{s} = \frac{B}{2} \gamma^{2} +  \frac{\kappa}{2} H^{2} + \kappa_{G} K,
\end{equation}
where $ \gamma  $ is the strain due to compression or
dilation of the smectic layers, $ H $ and $ K $ respectively denote the
mean curvature and the Gaussian curvature of the smectic layers
~\cite{deGProst}.

Finally 
\begin{equation}
\label{eqn:fc}
f_{c}   =  
\frac{K_{A}}{2} D_{i} c^{j} D^{i} c_{j}
   + \frac{K_{B}}{2} ( (N^{\alpha} \partial_{\alpha}) {\bf c} )^{2},
\end{equation}   
accounts for the energy cost for distortions in the $ {\bf c} $- field
which do not  arise  from distortions in the Frank director $ {\bf n} $
but from distortions in the smectic layers.
In ({\ref{eqn:fc}), 
$ D_{i} c^{j} = \partial_{i} c^{j} + \, ^{(2)}\gamma^{i}_{k} A_{i} c^{k}
$, 
$ A_{i} $ are components of the spin connection defined below, 
$ ^{(2)}\gamma^{i}_{k} $ is the completely antisymmetric unit tensor for
the two-dimensional smectic layers,  $ i,j,k = 1,2 $ refer to the internal
coordinates of the  smectic layers, and $ \alpha = 1,2,3 $. The spin
connection defined via $ K = \,  ^{(2)}\gamma^{ij} \partial_{i} A_{j} $ 
represents the frustration induced in the $ {\bf c} $- field on layers
with nonzero Gaussian curvature . The term with the coefficient $ K_{A} $ in 
(\ref{eqn:fc}) can be schematically written as 
$ (K_{A}/2)(S - K) \frac{1}{- \nabla^{2} } (S - K) $, 
where $ S $ is the disclination density and $ \nabla^{2} $ is the
covariant Laplacian in the two- metric $ ^{(2)}g_{ij} $ describing the
geometry of the smectic layers. {\em Gaussian curvature screens 
disclination charges} ~\cite{DN,PL}. Stability requires that  
$ B, \kappa, K_{A}, K_{B} > 0 $. 
Since Gaussian curvature is a total derivative, the coefficient 
$ \kappa_{G} $ can either be positive or negative. 
Thus the full free energy for our model is
$ F = \int ( f_{L} + f_{n} + f_{s} + f_{c}) dV $ ~\cite{MTW}.

If the coefficient of the Gaussian curvature term in
(\ref{eqn:fs}) is {\em negative}, a  double-tilt cylindrical
structure of radius $ R $  with a $ +1 $  disclination in the $ {\bf c} $-
field {\em lowers} the energy. However the smectic layers in this conical
configuration are bent. We now show that the free energy cost from mean 
curvature, $ {\bf c} $- field distortion, and surface tension can be 
compensated by the free energy gain from Gaussian curvature to stabilize 
a lattice of disclination cones (Fig.~\ref{conephase}).
 
Let us consider a triangular lattice of cylindrical stacks of cones 
(i.e., a triangular lattice of straight +1 disclination lines), 
and fill in the gaps between cylinders with nematic material such that the 
nematic director is parallel to the disclination cores. Such a configuration
minimizes the interfacial energy between the nematic and the smectic.
In what follows we ignore the repulsive interaction energy 
between the disclination lines. In the continuum, a conical stack of
smectic layers with its core along the z-axis can be parametrized by a 
position vector $ {\bf R} = (r \cos \phi, r \sin \phi, m r + z ) $ in
the cylindrical polar coordinate system. We choose $ m = \tan \theta $, 
where $ \theta $ is the tilt angle of the nematic director with respect 
to the local layer normal and $ D = d/\cos \theta $, where $ d $ is the 
equilibrium smectic layer spacing. This choice ensures that there is no 
distortion in the nematic director, and that the inter-layer separation 
along the local layer normal equals $ d $. The disclination density is a 
delta function, $ S = 2 \pi \delta( r, \phi ) / \sqrt{g} $, where $ g $
is the determinant of the two-metric. 
The Gaussian curvature is zero everywhere except on the $ z $- axis with 
a line curvature charge 
$ \Sigma_{+} $: $ K = 2 \pi \Sigma_{+} \delta(r, \phi)/\sqrt{g} $, 
with $ \Sigma_{+} = 1 - \frac{1}{ \sqrt{ 1 + m^{2} } } $. 
For cones with radius $ R $ the energy per unit length (along the cone
axes) of the triangular lattice  with reference to the equilibrium
smectic-$C$ is 
\begin{eqnarray}
\label{eqnarray:conenergy}
 f_{cone} & = &  (\sqrt{3} - \frac{\pi}{2}) a (T_{c} - T) R^{2}
\nonumber \\
& & +  \frac{\pi}{2} \left(  \kappa \frac{m^{2}}{ \sqrt{1 + m^{2} } } 
+  K_{A} \frac{m^{2}}{(1 + m^{2})^{3/2} } \right) \ln \frac{R}{a}
\nonumber \\
& &   +  \pi \kappa_{G} \left( 1- \frac{1}{\sqrt{1 + m^{2} }} \right)      
      +  \pi R \sigma,
\end{eqnarray}           
where $ a $ is a cutoff length of molecular dimension, and we have
included the contribution from the interfacial tension $ \sigma $ between
smectic-$ C $ and nematic. The Frank free energy does not enter into
the energetics of the cone phase because the Frank director is free of
distortions. Although the Gaussian curvature term in 
(\ref{eqn:fs}) integrates to the boundary, it nevertheless contributes
to the energetics because the double-tilt cylinders have a finite 
radius. The smectic-$C$ - nematic interfaces at the boundaries of the
double-tilt cylinders are essential for the
stabilization of the cone phase. We also note that  the Gaussian curvature
term drops out of the energetics for $ m = 0 $. It is therefore
clear that double-tilt cylinders cannot be energetically favored
near a second-order  $NC$ transition in which $ m $ grows continuously
from zero in the smectic-$C$ phase. In fact a relatively strong first
order transition with sufficiently small (large in magnitude and
negative in sign) value of $ \kappa_{G} $ would favor the formation of
stable double-tilt cylinders.

We assume the following reasonable values for the parameters entering
(\ref{eqnarray:conenergy}): $ \kappa \simeq 10^{-6} $ dyne, $ K_{A}
\simeq 3 \times 10^{-7} $ dyne, $ \sigma \simeq 10^{-2} $ dyne/cm
~\cite{Exp}. At the first order $ NC $ transition point there is no 
free energy difference between the two phases. With $ m = 1/2 $
~\cite{Exp1}, double-tilt cylinders of radius $ R \leq 50  \AA $      
are energetically favored over a distortion free smectic-$C$ provided $
\kappa_{G} \leq -3 \times 10^{-6} $ dyne. We are not aware of any
measurements of $ \kappa_{G} $ in thermotropic liquid crystals. However,
some lyotropic liquid crystals have $ \kappa_{G} $ values of this sign and
magnitude ~\cite{Exp2}. It is clear that the energy of a double-tilt
cylinder is lowered for smaller values of $ R $. It is interesting to
contrast this with the case of blue phases of cubic symmetry where
the pitch of the helix provides a {\it natural} length scale which
determines the radius of double-twist cylinders. There is no such natural
length scale determining the radius of double-tilt cylinders of the cone
phase ~\cite{footnote}. We therefore expect the radius of the double-tilt
cylinders to be of the order of a few times the cutoff length, which is
the smallest possible radius consistent with the conical structure.

Purely from a molecular point of view the tilt of the molecular director
in the smectic-$C$ owes its origin to the fact that a mutual
displacement of the molecules along their long axes lowers the 
interaction energy ~\cite{Madhusudana}. If {\it all} the neighbors
of a given molecule are displaced relative to it by an equal amount in the
same direction, the interaction energy can be lowered further. This is 
precisely the molecular configuration near the apex of the cones forming
double-tilt cylinders. This tendency of the molecules may contribute
significantly towards rendering $ \kappa_{G} $ negative.
    
Narrow double-tilt cylinders arrange themselves in a triangular
lattice to form the cone phase. With a decrease in temperature from the
transition point, the energy cost for the nematic which fills
the gap between the cylinders would become prohibitively
large. Consequently there would be a transition from the cone phase 
to the smectic-$C$ phase. We expect the temperature range of the cone 
phase to be narrow (of the order  $ 1^{0} $ C) as in the case of the
blue phases. To our knowledge such a phase has not been reported in the
literature. However there are several examples of compounds which exhibit
a {\it skew-cybotactic} short-range order over a wide (about $ 50^{0} $ C)
temperature range ~\cite{Exp3,Exp4}. X-ray and neutron
scattering experiments show that these systems exhibit several orders of
scattering maxima corresponding to molecular layering {\it along} the
nematic director in addition to the scattering due to tilt-order in the
layers. This implies that the molecular positions are correlated along the
director. This scattering pattern corresponding to fiber-like structures
can be interpreted as arising from the narrow double-tilt cylinders
proposed above. Since the medium is in the nematic phase, there is no
long-range positional order of the double-tilt cylinders. Indeed, these
`nematics' may in fact be in a {\it molten line liquid  phase} which has been
discussed in the context of directed polymers in  nematics, and flux
lattices ~\cite{Nelson}. As is the case for flux tubes in
superconductors, double-tilt disclination lines have a
purely repulsive logarithmic interaction potential. The melting of 
the cone phase should therefore be analogous to the melting of the
Abrikosov flux lattice, where the `braiding entropy' from the 
entanglement of the flux lines wins over the potential energy cost
at the melting point. Detailed analysis of the melting of the cone
phase into the line liquid phase is outside the scope of this letter.
We note that the foregoing analysis is also applicable to 
structures such as the smectic-$I$ which has hexatic order with 
molecular tilt oriented along bonds connecting nearest neighbors
~\cite{deGProst}. 

Further experiments to confirm the work presented in this letter  are
clearly of great interest. We have taken up some experiments to test these
ideas in our laboratory.

We thank V. A. Raghunathan, Madan Rao and Joseph Samuel for useful
discussions.

\vbox{
\vspace{0.5cm}
\epsfxsize=6.0cm
\epsfysize=6.5cm
\epsffile{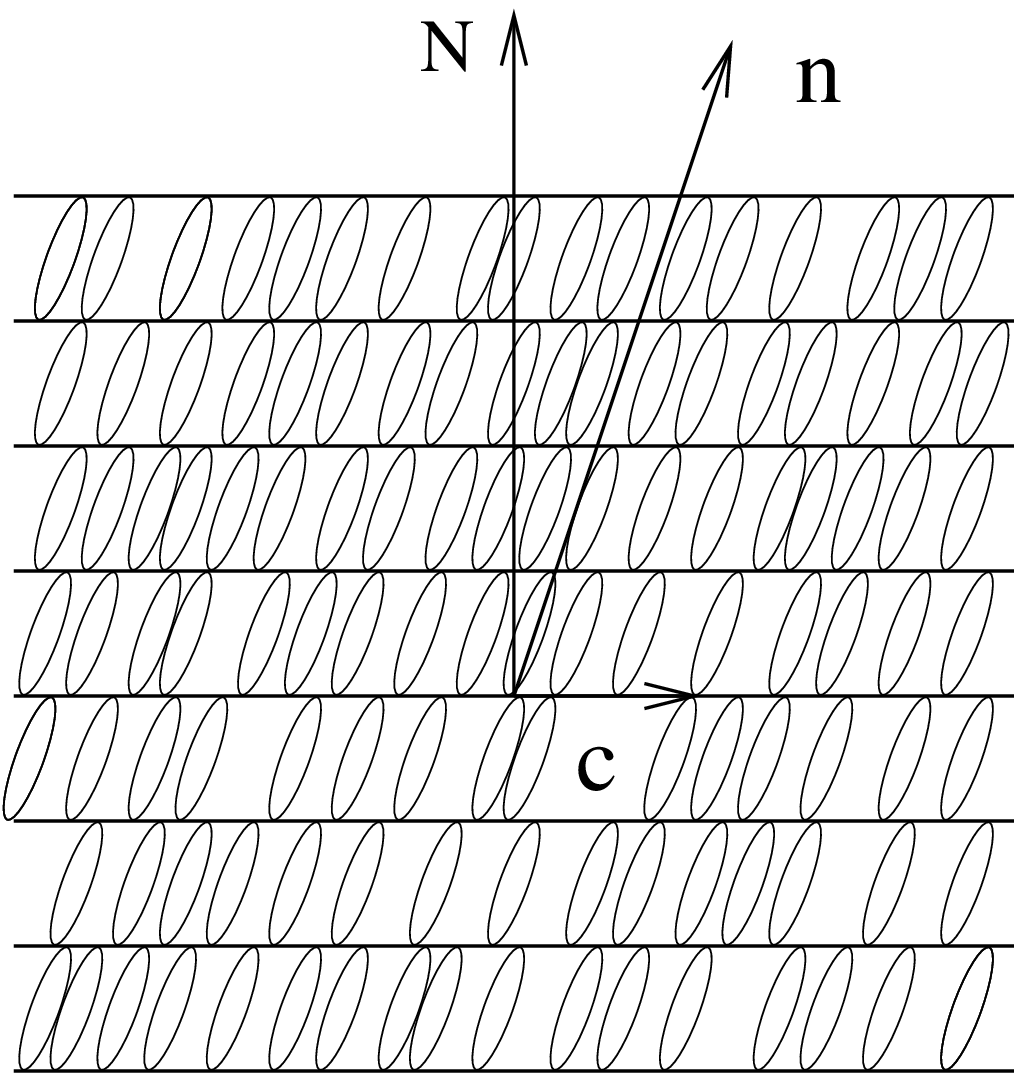}
\begin{figure}
\caption{ Schematic of the equilibrium smectic-$C$ structure. 
\label{smc}}
\end{figure}}
\vbox{
\epsfxsize=6.0cm
\epsfysize=6.5cm
\epsffile{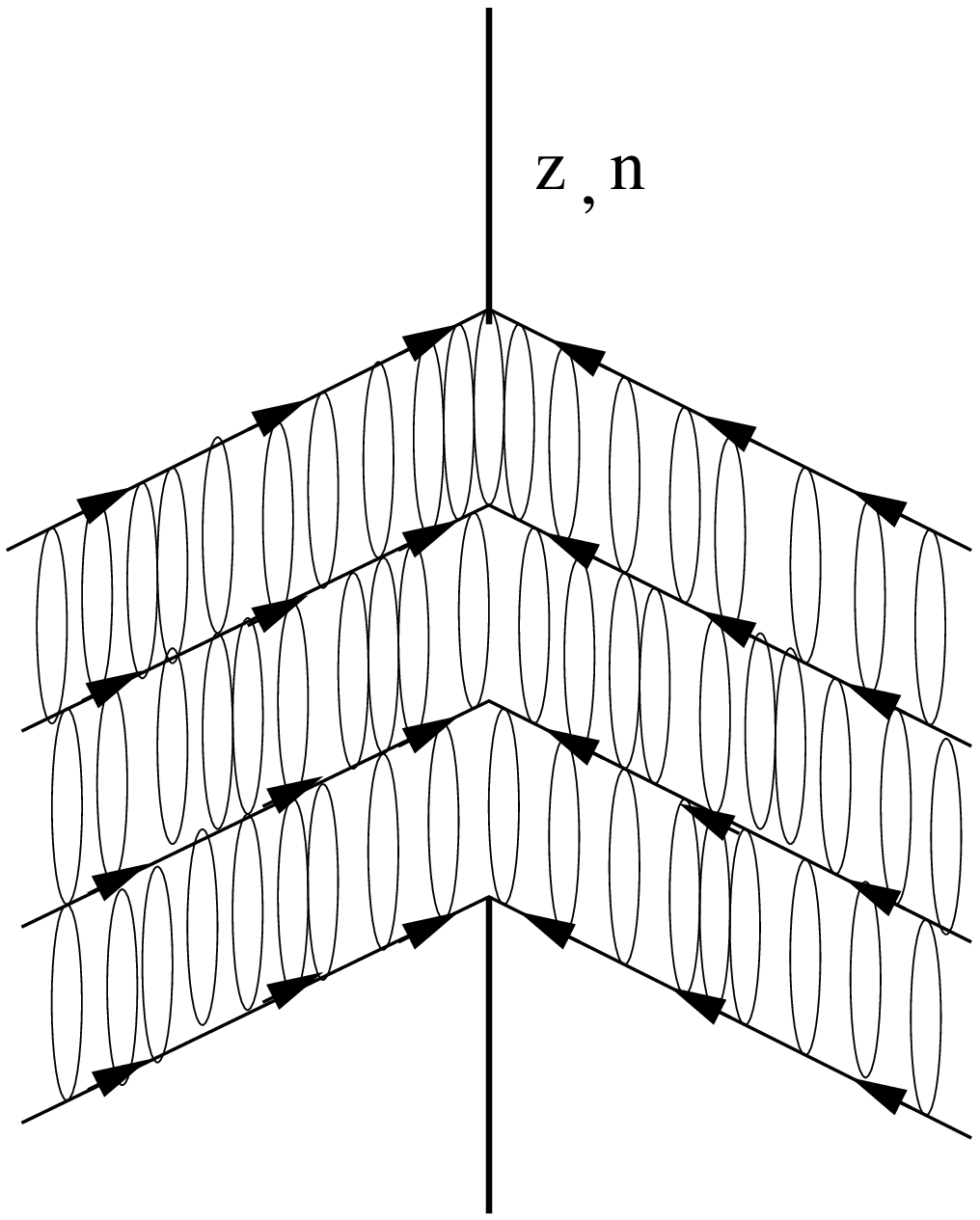}
\begin{figure}
\caption{ Schematic of a double-tilt cylinder. The nematic director
$ {\bf n} $ is parallel to the cone axis, which is a +1 disclination
line. The ${\bf c}$- vector (shown by the arrows in the figure) lies in
the smectic layers and points towards the cone axis. Note that the $ {\bf
n} $- field has no distortion, whereas the smectic layers bend. There is
no change in the smectic layer spacing along the local layer normal. 
\label{cylinder}}
\end{figure}}
\vbox{
\epsfxsize=7.0cm
\epsfysize=8.0cm
\epsffile{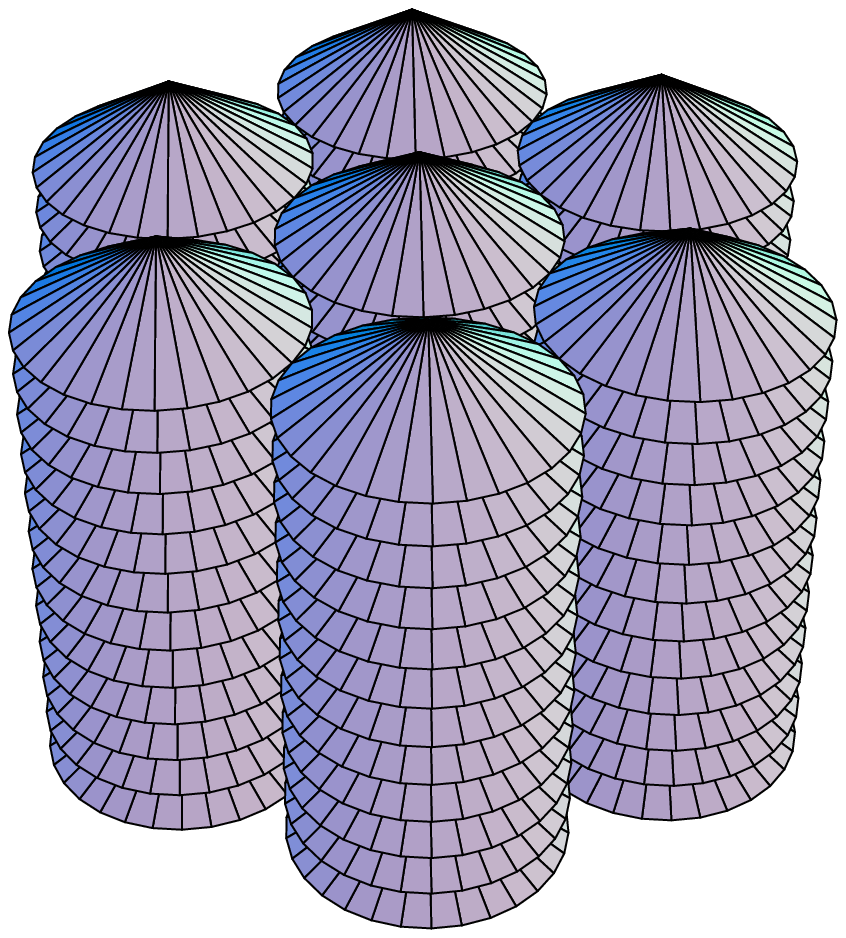}
\begin{figure}
\caption{ Schematic of the cone phase - a triangular lattice of
double-tilt cylinders. Each cylindrical stack of cones is a 
double-tilt cylinder depicted in (Fig.~\ref{cylinder}).
\label{conephase}}
\end{figure}}


\begin{references}
\bibitem{footnote} We note that this is strictly correct only for the
model discussed in this letter, where we have ignored higher order
terms in the elastic free energy.
\end{references}
\end{document}